\documentclass[journal=jctcce,manuscript=article]{achemso}
\SectionNumbersOn
\usepackage[version=4]{mhchem} % Formula subscripts using \ce{}
\setkeys{acs}{maxauthors = 100}

\usepackage{amsmath}
\usepackage{amssymb}
\usepackage{pdfpages}
\usepackage{url}
\usepackage{hyperref}

\author{Madlen Maria Reiner}
\affiliation{Faculty of Physics, University of Vienna, 1090 Vienna, Austria}
\alsoaffiliation{Vienna Doctoral School in Physics, University of Vienna, 1090 Vienna, Austria}
\alsoaffiliation{Institute of Theoretical Chemistry, Faculty of Chemistry, University of Vienna, 1090 Vienna, Austria}
\author{Johannes C. B. Dietschreit}
\email{johannes.dietschreit@univie.ac.at}
\affiliation{Institute of Theoretical Chemistry, Faculty of Chemistry, University of Vienna, 1090 Vienna, Austria}
\author{Leticia González}
\affiliation{Institute of Theoretical Chemistry, Faculty of Chemistry, University of Vienna, 1090 Vienna, Austria}
\email{leticia.gonzalez@univie.ac.at}
\author{Christoph Dellago}
\email{christoph.dellago@univie.ac.at}
\affiliation{Faculty of Physics, University of Vienna, 1090 Vienna, Austria}

\title{Nonadiabatic Forward-Flux Sampling of Rare Molecular Gas--Phase Ammonia Photodissociation}

\abbreviations{PES, NNP}

\begin{document}

\makeatletter
\setlength\acs@tocentry@height{4.45cm}
\setlength\acs@tocentry@width{8.25cm}
\makeatother
\begin{tocentry}
\centering
\includegraphics[width=8cm]{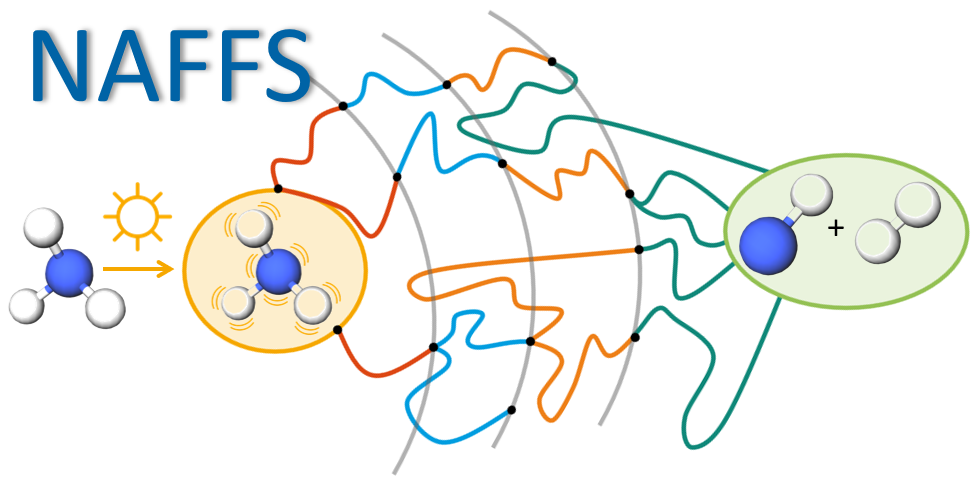}
\end{tocentry}

\begin{abstract}
    Simulating rare events in photochemistry by brute-force surface hopping is computationally prohibitive: most propagated trajectories either remain nonreactive or follow dominant relaxation channels, while the reaction of interest may occur with very low probability.
    Path sampling methods have long addressed this problem for ground-state dynamics, but their extension to nonadiabatic processes has so far been limited to low-dimensional analytical models.
    Here, we demonstrate that our recently developed nonadiabatic forward-flux sampling (NAFFS) method can be applied to efficiently simulate rare-events in full-dimensional molecular systems.
    As a representative benchmark, we investigate the rare molecular photodissociation channel of gas-phase ammonia, \ce{NH3 + h\nu \rightarrow NH + H2}. 
    NAFFS samples reactive trajectories with the correct statistical weight, reproduces the reaction rate constant obtained in previous brute-force dynamical studies, provides direct access to mechanistic information through an averaged committor analysis, and reduces the simulation time per transition trajectory by up to three orders of magnitude. These results establish NAFFS as an efficient and statistically rigorous framework for investigating rare nonadiabatic processes in realistic molecular systems.
\end{abstract}

\section{Introduction}
In molecular dynamics simulations, the rare event problem arises from a separation of time scales between the integration time step required to accurately describe the nuclear motion and the characteristic time scale on which the event of interest occurs, with the latter often exceeding the former by several orders of magnitude.\cite{Dellago2009}
    Conceptually, a rare event corresponds to a transition between two long-lived states separated by a substantial barrier, the crossing of which constitutes the event of interest.
    Brute-force simulations are inefficient in this regime, since most of the computational effort is spent on propagating the system within one of the metastable basins rather than on observing the event itself.
    The problem is further aggravated when the rare event involves several electronically excited states, so that a ground-state treatment\cite{Pigeon2026,Dietschreit2022a,Bocus2026arXiv} is no longer sufficient and computationally expensive nonadiabatic dynamics needs to be employed.
    
    % method
    Our recently proposed nonadiabatic forward-flux sampling (NAFFS) methodology\cite{Reiner2023} addresses the challenge of sampling such rare events in photochemical reactions---in particular, those associated with high barriers on excited-state potential energy surfaces (PESs) or with small nonadiabatic or spin--orbit couplings between electronic states.
    By selectively generating trajectories along the reaction channel of interest, NAFFS avoids the inherent inefficiency of brute-force dynamical simulations, in which the unconstrained propagation of the system yields only a small fraction of trajectories relevant to the reaction under study.
    This idea underpins all of the well-established ground-state path sampling methods,\cite{Dellago1998a} such as transition interface sampling\cite{VanErp2003} and forward-flux sampling.\cite{Allen2005,Allen2006,Allen2006a}
    NAFFS extends forward-flux sampling to nonadiabatic surface hopping dynamics,\cite{Tully1971} has been implemented in the surface hopping including arbitrary couplings (SHARC) framework,\cite{Richter2011,Mai2018} and, to date,  has been validated only for chemically motivated analytical model systems.
    
    Here, we apply NAFFS to the rare molecular photodissociation channel of gas-phase ammonia,
    \begin{equation}
        \mathrm{NH_3} \overset{h\nu}{\longrightarrow} \mathrm{NH+H_2}
        \label{eq:molecular_channel}
    \end{equation}
    thereby providing the first full-dimensional molecular application. 

    By selectively sampling the reactive trajectories of interest, NAFFS avoids expending computational effort on propagating the dominant pathways until the statistics for the rare channel converge. 
    In particular---reflecting a characteristic of path sampling techniques---this is done without introducing biasing potentials, as employed in other rare event sampling approaches.\cite{Kamenik2021,Liu2025,Henin2022} 
    To our knowledge, this is the first application of a nonadiabatic path sampling method to a realistic molecular system, moving beyond proof-of-concept model studies.\cite{Sherman2016,Schile2018,Ghamari2026,Yang2026}
    This demonstrates the scalability of NAFFS from low-dimensional analytical models to full-dimensional molecular systems, providing an efficient framework to investigate rare photochemical events beyond the practical limits of brute-force nonadiabatic dynamics.

    % ammonia
    Ammonia provides a particularly suitable molecular test case for NAFFS because its photodissociation dynamics involves several competing product channels with strongly unequal probabilities.
    Upon absorption of a photon, ammonia is promoted to an excited electronic state, from which several photodissociation pathways become accessible:  a dominant radical channel (\ce{NH3 + h\nu \rightarrow NH2 + H}) and a rare molecular one (\ce{NH3 + h\nu \rightarrow NH + H2}). 
    Although the photochemistry of ammonia has been studied extensively,\cite{Biesner1988,Biesner1989,Woodbridge1991,Dixon1996,Bach2003,Nangia2006,Hause2006,Li2007,Lai2008,Chatterley2013,Rodriguez2014,Xie2015,Guan2020,Han2020,Wang2021b,Zhao2023} the rare molecular photodissociation channel has received comparatively less attention, both experimentally\cite{Mordaunt1996,Sato2001,Leach2005} and computationally.\cite{Runau1977,Xie2014,Wang2022,Bachmair2025}
    Recent surface-hopping simulations\cite{Wang2022} based on neural-network potentials (NNPs)\cite{Wang2021a} of the two lowest singlet and the lowest triplet states, found that the molecular channel is reached by only about 2\% of the propagated trajectories, compared to roughly 97\% for the radical channel.
    
    A follow-up study\cite{Bachmair2025} using modified PESs reported that 94\% of the trajectories yield \ce{NH2 + H}, while only 0.6\% follow the rare molecular channel.
    The combination of a very low branching ratio and the involvement of several coupled electronic excited states, makes the molecular channel a stringent benchmark for assessing the performance of NAFFS.

    The remainder of the paper is structured as follows. 
    Section~\ref{sec:methods} summarizes the theory behind NAFFS and describes the computational methods used, how they are combined, and the details of their implementation.
    Section~\ref{sec:results} presents and discusses the results, and Section~\ref{sec:conclusions} provides our conclusions.

%%%%%%%%%%%%%%%%%%%%%%%%%%%%%%%%%%%%%%%%%%%%%%%%%%%%%%%%%%%%%%%%%%%%%%%%%%%%%%%%%%%%%%%%%%%%%
%%%%%%%%%%%%%%%%%%%%%%%%%%%%%%%%%%%%%%%%%%%%%%%%%%%%%%%%%%%%%%%%%%%%%%%%%%%%%%%%%%%%%%%%%%%%%
    \section{Methods}
    \label{sec:methods}

    %%%%%%%%%%%%%%%%%%%%%%%%%%%%%%%%%%%%%%%%%%%%%%%%%%%%%%%%%%%%%%%%%%%%%%%%%%%%%%%%%%%%%%%%%%%%%
    \subsection{Theory}
    \label{sec:theory}

    Surface hopping is a mixed quantum--classical simulation technique in which the nuclei are treated as classical particles and nonadiabatic effects are accounted for through stochastic transitions between multiple electronic states.\cite{Mai2020b,Nelson2020,Barbatti2011}
    The nuclei are propagated according to Newton's equations of motion on a single active PES,  while the electronic wave function evolves quantum mechanically in a basis of coupled electronic states.
    Stochastic hops between electronic states are driven by nonadiabatic and spin--orbit couplings, with hopping probabilities chosen to reproduce the quantum electronic populations while minimizing the number of state transitions (\textit{i.e.}, fewest switches).
    Decoherence corrections are commonly applied to suppress unphysical electronic coherence following surface hops.\cite{Crespo-Otero2018,Nelson2020,Mai2020b,Heindl2021}

    Building on surface hopping, the NAFFS methodology\cite{Reiner2023} enables efficient sampling of rare nonadiabatic photochemical events involving multiple PESs. 

    In a nutshell, for a given photoreaction containing a rare event, one defines an initial reactant region $A$ and a final product region $B$, together with a set of $n$ interfaces between them, expressed in terms of collective variables, for example distances, angles, dihedrals or coordination numbers. 
    The definitions of these regions and interfaces may also include or exclude specific electronic states.
    
\begin{figure}
    	\includegraphics[width=0.8\textwidth]{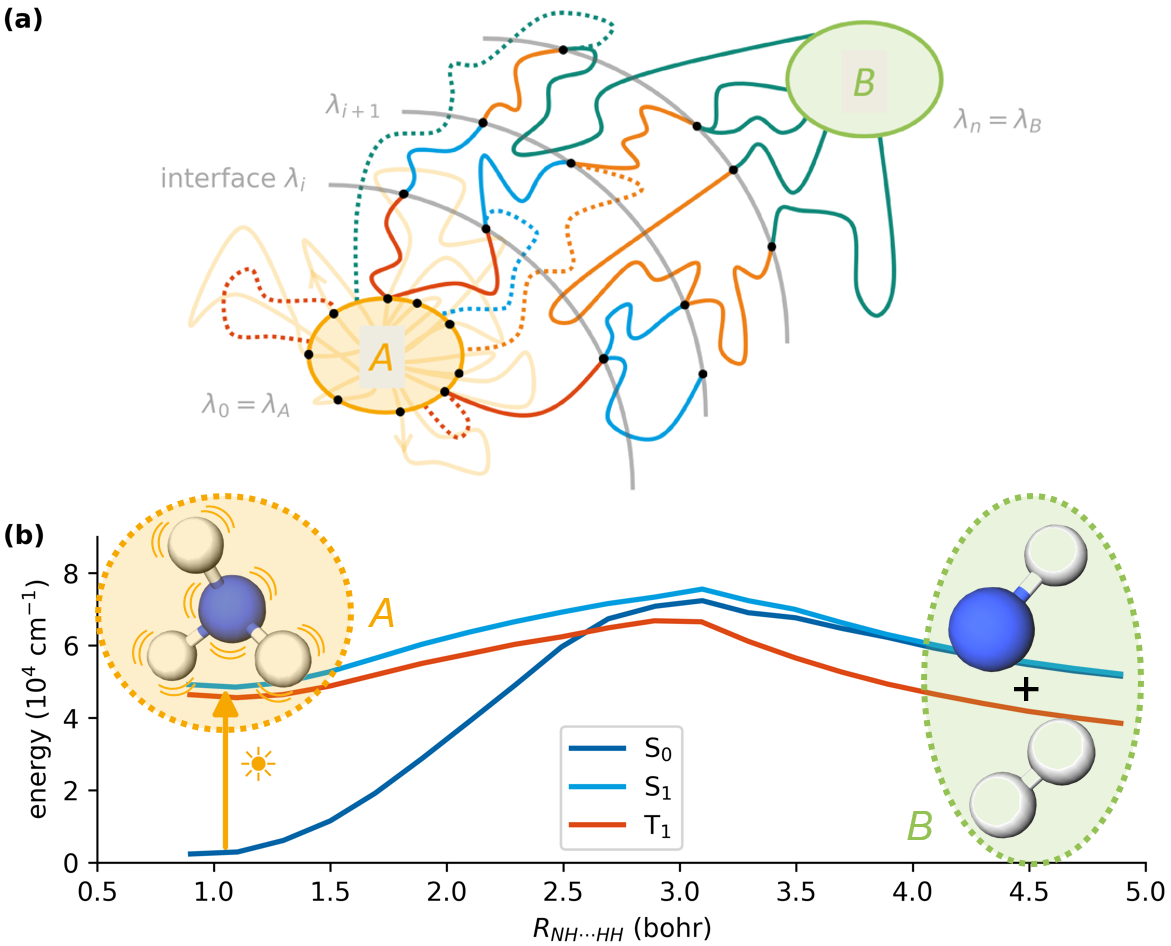}
    	\caption{a) Schematic depiction of the forward-flux sampling method. 
        The initial region $A$ is shown in yellow, the final region $B$ in green, and interfaces in gray. 
        The flux simulation is shown as light yellow solid line. 
        Shooting points are depicted as black dots at the boundary of $A$ (flux events) and at the interfaces. 
        Interface-crossing trajectories are shown in red, blue, orange, and dark green as solid lines. 
        Trajectories returning to the initial region $A$ are drawn as dotted lines. 
        b) Lowest two singlet (blue solid lines) and lowest triplet (red solid line) potential energy curves along an \ce{NH + H2} linear synchronous transit dissociation path\cite{Wang2021a}. $R_\mathrm{NH-HH}$ denotes the center of mass distance between \ce{NH} and \ce{H2}.
        }
    	\label{fig:theory}
    \end{figure}

    NAFFS then samples surface hopping transition trajectories connecting $A$ and $B$ in a multi-step procedure (see Fig.~\ref{fig:theory}a).
    First, a \textit{flux simulation} is performed, consisting of a surface hopping %FSSH 
    simulation initiated in $A$. 
    Whenever the system leaves $A$, the corresponding configuration is stored as an \textit{initial shooting point}, also called \textit{flux event}.
    Second, starting from randomly chosen initial shooting points, short surface hopping molecular dynamics trajectories, referred to as \textit{shots}, are launched by continuing the dynamics from those points.
    Because the dynamics is stochastic, as required by any FFS-type method, two trajectories initiated from the same shooting point are not identical.
    Each shot is either accepted, if it reaches the next interface---its endpoint is then stored as a new shooting point for the subsequent NAFFS cycle---or rejected, if it returns to $A$. 
    We call the accepted trajectories {\em interface-crossing} trajectories. 
    This procedure is repeated at each interface until the final product region $B$ is reached.
    The partial trajectories landing in $B$ in the last step are concatenated with the corresponding interface-crossing partial trajectories from preceding interfaces to yield the final transition trajectories from $A$ to $B$.
    Importantly, because NAFFS is a path sampling method, the underlying dynamics remains entirely unbiased, and the resulting ensemble of transition trajectories is the same as that obtained in an unconstrained brute-force dynamics simulation.\cite{Dellago2009}

    The rate constant $k_{AB}$ of the reaction from $A$ to $B$ is given by
    \begin{equation}
        k_{AB} = \phi_A \cdot \prod\limits_{i=0}^n P_A (\lambda_{i+1}\vert \lambda_i)\ ,
        \label{eq:rate}
    \end{equation}
    where $\phi_A$ is the flux out of the initial region $A$, which is calculated by dividing the number of times that $A$ is left during the flux simulation by the total length of the flux simulation. 
    $P_A(\lambda_{i+1}\vert \lambda_i)$ is the conditional probability that a shot initiated at interface $\lambda_{i}$ reaches interface $\lambda_{i+1}$ before returning to $A$. 
    Here and in the following, each interface is denoted by the value $\lambda_i$ of the collective variable $\lambda$ that defines it. 
    Operationally, the conditional probability $P_A(\lambda_{i+1}\vert \lambda_i)$, also called the {\em crossing probability}, is estimated from the number $N_{i+1}$ of partial trajectories initiated at interface $\lambda_{i}$ that reach the next interface $\lambda_{i+1}$ divided by the total number of shots $M_i$ from interface $\lambda_{i}$,
    \begin{equation}
        P_A(\lambda_{i+1}\vert \lambda_i) = \dfrac{N_{i+1}}{M_i}
        \label{eq:acceptance_prob}
    \end{equation}
    There are $n+2$ interfaces in total, with $\lambda_0=\lambda_A$ defining the boundary of $A$ and $\lambda_{n+1}=\lambda_B$ defining the boundary of $B$.
    The absolute error of the rate constant $\Delta k_{AB}$ can be estimated  as\cite{Allen2006a,Hussain2020,Borrero2008}
    \begin{equation}
        \Delta k_{AB}=k_{AB}\sqrt{\frac{1}{N_0}+\sum\limits_{i=0}^{n}\frac{1-P_A(\lambda_{i+1}|\lambda_i)}{P_A(\lambda_{i+1}|\lambda_i)\cdot M_i}}
        \label{eq:FFSerror}
    \end{equation}
    where $N_0$ is the number of flux events. 
    Further details on the error estimation can be found in the original reference \citenum{Reiner2023}.

    \subsection{Implementation}
    \label{sec:implementation}
    
    Our NAFFS implementation\cite{Reiner2023} couples the Open Path Sampling (OPS) package\cite{Swenson2019,Swenson2019a} to the SHARC dynamics library\cite{Mai2019,Plasser2019,Plasser2019a}.
    SHARC implements the trajectory surface hopping method of the same name\cite{Richter2011,Mai2015,Mai2018} for the semi-classical propagation of the nuclear coordinates.
 
    In this work, we employ SHARC~4.0.\cite{Mai2025,Mausenberger2025}
    Owing to the modular design between OPS and SHARC, combining NAFFS with the NNP for the ammonia molecule\cite{Wang2021a} is straightforward.
    More generally, any method providing energies, forces, and couplings available through a SHARC interface---such as analytical potentials, linear-vibronic coupling models, machine learning potentials, or quantum chemistry programs---is also directly accessible to NAFFS. 
    This gives users the flexibility to perform NAFFS simulations with a broad range of electronic-structure methods.
    
    To ensure transparency and reproducibility, all data and code used in this work are freely available (see Sec.~S1).

    %%%%%%%%%%%%%%%%%%%%%%%%%%%%%%%%%%%%%%%%%%%%%%%%%%%%%%%%%%%%%%%%%%%%%%%%%%%%%%%%%%%%%%%%%%%%%
    \subsection{System}
    \label{sec:system}
    
    As a first application of the NAFFS algorithm to a real chemical system, we consider the rare molecular photodissociation of gas-phase ammonia, Eq.~(\ref{eq:molecular_channel}).\cite{Runau1977,Xie2014,Wang2022}
    The corresponding PESs and electronic-structure quantities are taken from the recently published NNP representation of Wang \textit{et al.},\cite{Wang2021a} in the form used in our recent work.\cite{Bachmair2024,Bachmair2025}
    The NNP describes the two lowest singlet states (S$_0$, S$_1$) and the lowest triplet state (T$_1$) of ammonia, together with the corresponding forces, transition dipole moments, nonadiabatic couplings, and spin--orbit couplings.
    Such machine learning potentials provide an accurate surrogate for the underlying quantum chemical method at a small fraction of its cost, by fitting the relevant electronic-structure quantities on a training set of high-level reference data.\cite{Westermayr2020b,Westermayr2021,Westermayr2021a,Bachmair2022} The reference data were obtained from multi-reference configuration interaction calculations including single and double excitations (MRCISD) using molecular orbitals taken from a state-averaged multi-configurational self-consistent field calculation (SA-MCSCF), averaged over the three states with equal weights, with the aug-cc-pVTZ basis set, augmented by a Rydberg function on the nitrogen atom; full details of the representation are given in Wang \textit{et al.}\cite{Wang2021a}.

    %%%%%%%%%%%%%%%%%%%%%%%%%%%%%%%%%%%%%%%%%%%%%%%%%%%%%%%%%%%%%%%%%%%%%%%%%%%%%%%%%%%%%%%%%%%%%
    
    %%%%%%%%%%%%%%%%%%%%%%%%%%%%%%%%%%%%%%%%%%%%%%%%%%%%%%%%%%%%%%%%%%%%%%%%%%%%%%%%%%%%%%%%%%%%%
    \subsection{Computational Details}
    \label{sec:computational}

    The workflow for studying the rare, molecular photodissociation of \ce{NH3} (Eq.~(\ref{eq:molecular_channel})) with NAFFS is summarized in Fig.~\ref{fig:theory}.
    Configurations sampled around the ground-state minimum are vertically excited into the initial region $A$ (excited \ce{NH3}, see Fig.~\ref{fig:theory}b), from which NAFFS samples surface hopping trajectories reaching the final region $B$ (\ce{NH + H2}), while explicitly including the S$_0$, S$_1$ and T$_1$ electronic states.
    
    NAFFS supports two preprocessing steps that are optional and may be used individually or in combination.\cite{Reiner2023} 
    The first is the generation of initial conditions, either by creating ground-state configurations that are vertically excited to the relevant excited electronic states, from which the surface hopping dynamics originate, or by using an explicit laser field for the excitation. 
    The second is the equilibration into the reactant region $A$. 
    This step is necessary if the region populated immediately after excitation does not coincide with $A$, so that the system first needs to relax to the initial region before the NAFFS simulation can be started from there.
    
    For the first step, we use the excited initial conditions of Bachmair \textit{et al.},\cite{Bachmair2025} obtained by Wigner-sampling around the S$_0$ minimum, followed by the selection of configurations with the highest photoexcitation probability based on oscillator strengths and excitation energies using pySHARC.\cite{Plasser2019}
    The second, equilibration step is not needed in this case, because $A$ is defined in the geometric region of the S$_0$ minimum but only includes configurations in the excited state, so Wigner-sampled configurations excited to S$_1$ already lie inside $A$ (see also Sec.~S2).
    The vertically excited configurations are then used as starting points for the flux simulation, and the subsequent NAFFS cycles proceed as described in Sec.~\ref{sec:theory}.
    Region and interface definitions, together with the remaining simulation parameters, are given below.

    To map the photodissociation, we use collective variables based on the interatomic distances of the molecule.
    The three \ce{N-H} distances, ordered by decreasing length, are denoted by $r_{l}^{\rm NH}$, $r_{m}^{\rm NH}$, and $r_{s}^{\rm NH}$, and the three \ce{H-H} distances are analogously denoted by $r_{l}^{\rm HH}$, $r_{m}^{\rm HH}$, and $r_{s}^{\rm HH}$.

    The initial region $A$ contains all configurations with $r_{m}^{\rm NH} < 2.5$\,a$_0$ (where a$_0$ denotes the Bohr radius) and all other bonds below 3\,a$_0$, such that $A$ consists of undissociated \ce{NH3} configurations.
    The final region $B$, adapted from Wang \textit{et al.},\cite{Wang2022} is defined by $r_{m}^{\rm NH} > 10$\,a$_0$ with $r_{s}^{\rm NH}, r_{s}^{\rm HH} < 3$\,a$_0$, such that it contains \ce{NH + H2} configurations.
    We placed three interfaces along $r_{m}^{\rm NH}$ at $[2.69, 2.94, 3.57]$\,a$_0$, subject to $r_{s}^{\rm NH}, r_{s}^{\rm HH} < 3$\,a$_0$, yielding reasonable crossing probabilities (see Sec.~\ref{sec:results}).

\begin{figure}
    \includegraphics[width=0.8\textwidth]{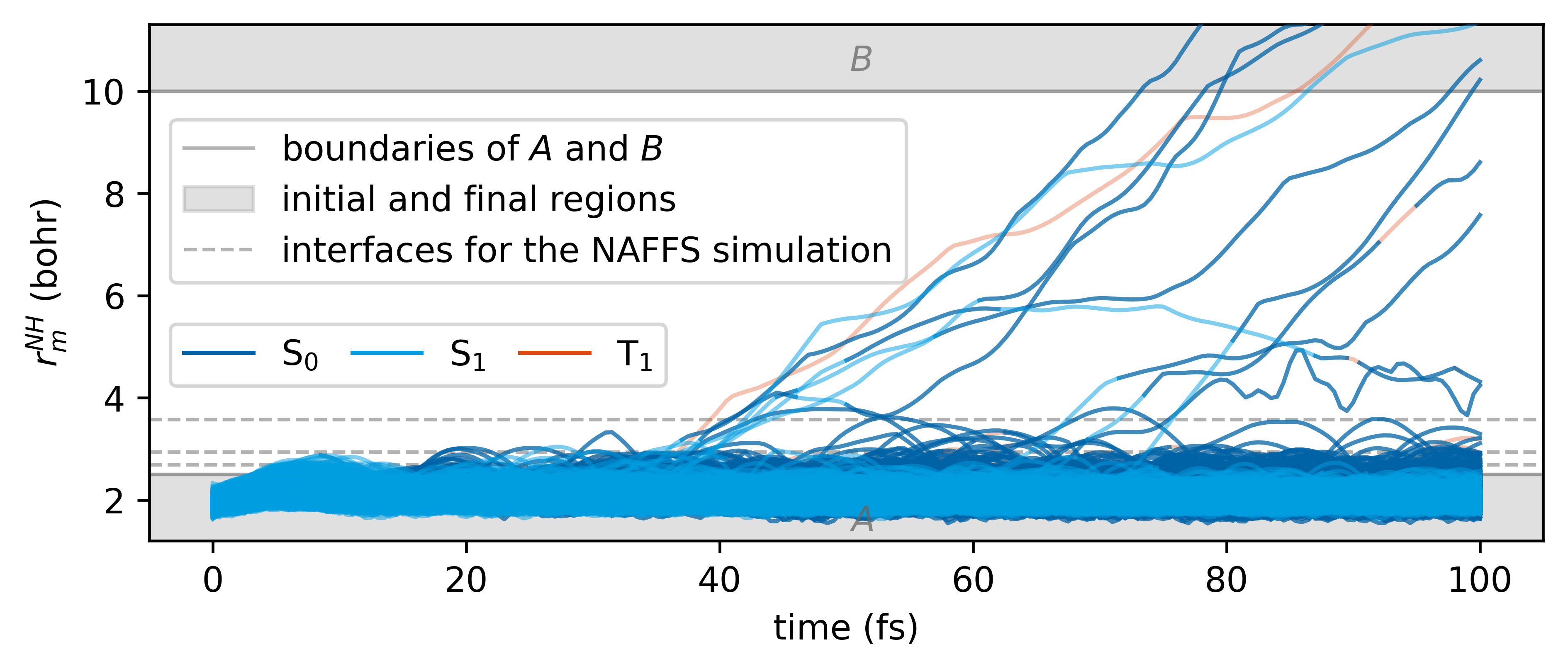}%\
    \caption{Flux trajectories as a function of the main geometric coordinate $r_{m}^{\rm NH}$. 
    The color of the trajectory marks the active electronic state (dark blue: S$_0$, light blue: S$_1$, red: T$_1$).
    Initial and final regions are highlighted in gray, interfaces as dashed lines. The starting points of the trajectories (time zero) are the initial S$_1$ configurations along $r_{m}^{\rm NH}$.}
    	\label{fig:inits_flux}
    \end{figure}

    The flux simulation has a total length of 500\,ps (5,000 initial conditions propagated for 100\,fs). 
    In the subsequent NAFFS cycles, we performed 1000 shots per interface.
    All trajectories are propagated with a time step of $0.5$~fs and 25 electronic substeps in the diagonal representation of SHARC.
    For frustrated hops, momenta are reflected along the nonadiabatic coupling vector; for accepted hops, the kinetic energy is rescaled along the same direction.
    Decoherence is treated with the energy-based correction using a standard decoherence parameter of $0.1$~Ha.\cite{Granucci2010}
    An example SHARC input file is provided in Sec.~S1.
    Figure~\ref{fig:inits_flux} shows the initial region $A$, which in addition to the geometric constraint only includes points on the electronic S$_1$ state, the final region $B$, the interfaces for the NAFFS simulation, and, as starting points of the shown flux trajectories, the initial S$_1$ configurations along $r_{m}^{\rm NH}$.

    The Langevin thermostat is parametrized with a friction constant $\gamma=16.79$~ps$^{-1}$ and a temperature $T=2375$~K, both derived from a set of thermostat-free brute-force trajectories.\cite{Bachmair2025}
    These reference trajectories start from 50,000 excited initial conditions propagated for up to 1~ps, with early termination upon reaction.
    From the 2305 non-dissociating trajectories of this set, we obtain our target temperature $T$ from the average kinetic energy.
    %,
    % \begin{equation}
    %     T = \frac{2 E_{\rm kin}}{(3N-6)k_{\rm B}}
    % \end{equation}
    % with Boltzmann's constant $k_{\rm B}$ and the number of atoms $N=4$.
    The friction constant is one order of magnitude lower than the slowest angular frequency present in the system, $\omega=180$~ps$^{-1}$ (oscillation period $T_{\rm osc}=35$~fs); the fastest frequency is $\omega=650$~ps$^{-1}$ ($T_{\rm osc}=10$~fs), allowing the thermostat to act more slowly than the system's natural vibrations, and hence, oscillations remain physical.
    These values are taken from experimental literature\cite{Johnson2022} and are consistent with the oscillation periods of $12$--$22$~fs observed in the non-dissociating reference trajectories.\cite{Bachmair2025}
    The dependence of the rate constant on the friction constant is discussed in Sec.~S3.

%%%%%%%%%%%%%%%%%%%%%%%%%%%%%%%%%%%%%%%%%%%%%%%%%%%%%%%%%%%%%%%%%%%%%%%%%%%%%%%%%%%%%%%%%%%%%
%%%%%%%%%%%%%%%%%%%%%%%%%%%%%%%%%%%%%%%%%%%%%%%%%%%%%%%%%%%%%%%%%%%%%%%%%%%%%%%%%%%%%%%%%%%%%
\section{Results and Discussion}
\label{sec:results}
    %discuss in more detail: statistical expectation vs our results 
    %%%%%%%%%%%%%%%%%%%%%%%%%%%%%%%%%%%%%%%%%%%%%%%%%%%%%%%%%%%%%%%%%%%%%%%%%%%%%%%%%%%%%%%%%%%%%
    \subsection{NAFFS simulation}
    %\label{sec:NAFFS}
    
    The flux simulations produced 736 initial shooting points on the boundary of $A$ for the first NAFFS cycle.
    Following Sec.~\ref{sec:theory}, the flux out of $A$ is
    \begin{equation}
        \phi_A = ( 1.426 \pm 0.054)~{\rm ps}^{-1} \ .
    \end{equation}
    The corresponding flux trajectories along the $r_m^{\ce{NH}}$ coordinate are shown in Fig.~\ref{fig:inits_flux}.
    
    The interface crossing probabilities are listed in Table~\ref{tab:acceptanceprobabilities}.
    All values lie between 20\% and 70\%, which is the regime typically targeted in FFS simulations: lower probabilities drastically reduce computational efficiency, while higher ones increase the correlation between adjacent interfaces and reduce the statistical reliability of the rate estimate.\cite{Kratzer2013}
    The chosen interface placement therefore represents a reasonable compromise between these two regimes.

    \begin{table}[tb]
        \centering
      \caption{Crossing probabilities $P_A(\lambda_{i+1}\vert \lambda_i)$ of the NAFFS interfaces $\lambda_i$. Note that $\lambda_0=\lambda_A$ and $\lambda_4=\lambda_B$.}
      \begin{tabular}{c c}
        \hline
        $i$ & $P_A(\lambda_{i+1}\vert \lambda_i)$ \\
        \hline
        0 & $(20.1 \pm 1.3)\%$ \\
        1 & $(22.2 \pm 1.4)\%$ \\
        2 & $(45.6 \pm 1.6)\%$ \\
        3 & $(64.6 \pm 1.6)\%$ \\\hline
      \end{tabular}
      \label{tab:acceptanceprobabilities}
    \end{table}

\begin{figure}
    	\includegraphics[width=0.8\textwidth]{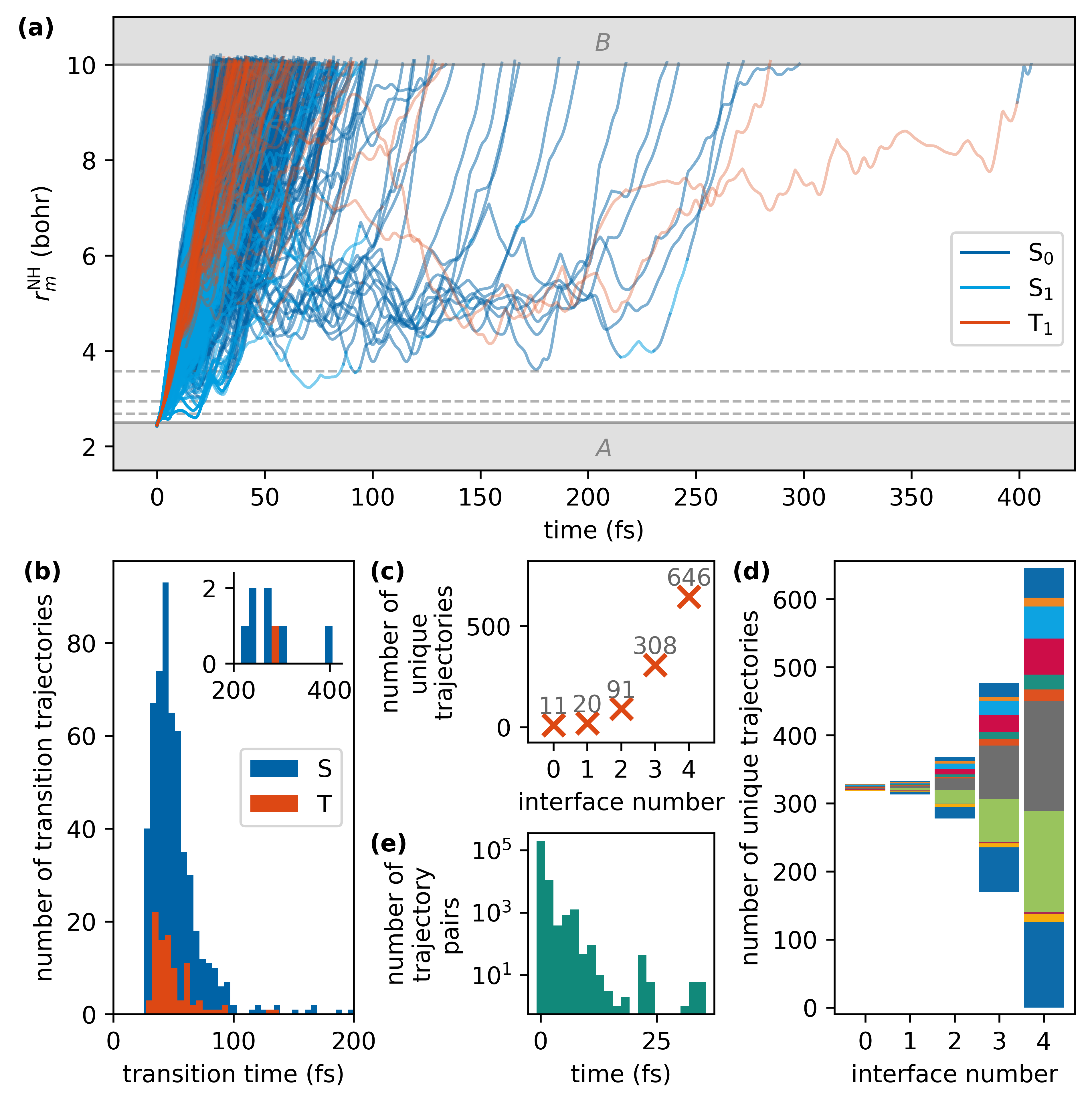}
    	\caption{a) Transition trajectories plotted along $r_{m}^{\rm NH}$ for the rare, molecular photodissociation channel of \ce{NH3}. 
        For each trajectory the point in time where the trajectory left the initial region $A$ is set to zero. 
        Initial and final regions are highlighted in gray, interfaces used in the simulation as dashed lines. 
        The color of the trajectory marks the electronic state (dark blue: S$_0$ state, light blue: S$_1$ state, red: T$_1$ state). 
        b) Distribution of transition times for the obtained reactive trajectories according to the state on which a each trajectory enters $B$ (S denotes singlet, T denotes triplet). The inlay shows the same distribution for transition times over 200~fs.
        c) Number of unique shooting points for each interface. 
        d) Visualization of the branching of the transition trajectories along the interfaces (see main text for detailed explanation).
        e) Distribution of shared starting sequences as a measure of correlation for all pairs of trajectories. Note that the $y$-scale is logarithmic for better visibility of counts for times larger than zero.
        }
    	\label{fig:transitiontrajectories}
    \end{figure}

    Combining the interface-crossing trajectories yields 646 reactive transition trajectories that form \ce{NH + H2}.
    These trajectories are shown in Fig.~\ref{fig:transitiontrajectories}a, projected onto $r_{m}^{\rm NH}$, with each trajectory shifted to a common time origin defined by its exit from $A$.
    The shortest reactive trajectory has a length of 25.5\,fs, the longest 406.0\,fs, and the average transition time is $(53.8 \pm 1.4)$\,fs, where the uncertainty denotes the standard error of the mean. 
    Figure~\ref{fig:transitiontrajectories}b shows the distribution of transition times, resolved by the spin multiplicity of the final configuration of each trajectory (\textit{i.e.}, whether the trajectory enters the final region $B$ on a singlet or a triplet state).
    Trajectories ending on a triplet state are, on average, shorter than those ending on a singlet state.
    Both distributions are heavy-tailed, with the singlet distribution showing a shoulder between roughly 70 and 100\,fs.
    Plots of the interface-crossing partial trajectories, together with convergence checks confirming that the size of the NAFFS simulation (the length of the flux simulation and the number of shots per interface) is sufficient, are provided in Sec.~S3.

    % UNIQUE FLUX EVENTS & FLUX EVENT ANALYSIS
    \subsection{Flux Events and Trajectory Correlation}
    \label{subsec:FluxEvents_TrajCorrelation}
    Because FFS is a branching method, partial trajectories accepted at interface $\lambda_i$ act as ancestors for multiple shots at $\lambda_{i+1}$, thus trajectories closer to $A$ share more common segments than those closer to $B$.
    Fig.~\ref{fig:transitiontrajectories}c quantifies this branching by showing the number of unique shooting points at each interface.
    The 646 final transition trajectories originate from only 11 unique flux events at the boundary of $A$. 
    Although this number might appear small, it is in good agreement with the statistical expectation: given the transition probabilities in Table~\ref{tab:acceptanceprobabilities} one would obtain 9.7 flux events leaving $A$ that reach $B$ from the 736 sampled flux events on the boundary of $A$.
    Figure~\ref{fig:transitiontrajectories}d visualizes the branching per unique flux event: each color represents one of the 11 unique flux events, and the width of each color block on a given interface represents the number of partial trajectories descending from it.
    The flux events contribute very unevenly to the final reactive ensemble, with the smallest contributing flux event giving rise to 12 transition trajectories and the largest to 162.
    
    If we compare the ``successful'' flux events, \textit{i.e.}, events from which the transition trajectories originate, with flux events producing trajectories discarded later, we see that they show overall smaller $r_l$ values on average (reporting means and standard errors of the mean, as well as the standard deviations (SD)):
    $r_l^{\ce{NH}}=(4.080\pm 0.012)$\,a$_0$ (SD: 0.30\,a$_0$) and 
    $r_l^{\ce{HH}}=(4.347\pm 0.031)$\,a$_0$ (SD: 0.79\,a$_0$) for the successful flux events \textit{vs.}
    $r_l^{\ce{NH}}=(6.684\pm 0.095)$\,a$_0$ (SD: 4.8\,a$_0$) and 
    $r_l^{\ce{HH}}=(7.433\pm 0.095)$\,a$_0$ (SD: 4.7\,a$_0$) for the unsuccessful flux events. 
    Since the other distances, $r_m$ and $r_s$, do not differ significantly between successful and unsuccessful flux events, the observed difference likely reflects an early stage of the dominant dissociation channel, in which a single \ce{H} atom dissociates and \ce{NH2} is formed. Such trajectories do not contribute to the targeted molecular dissociation channel and are therefore discarded.

    Counterintuitively, the small number of 11 shared flux events for reaction paths does not necessarily imply a strong correlation among the resulting transition trajectories.
    From a purely mathematical perspective, of the 208,335 possible pairs of trajectories---given by the binomial coefficient 646 over 2---only 0.8\% (1711 pairs from the binomial coefficient of $646/11\approx59$ over 2) share a common origin on average. 
    However, we need to consider not only the shared flux events, but also the total shared starting sequences. 
    Figure~\ref{fig:transitiontrajectories}e shows the distribution of the starting sequence lengths that the transition trajectories have in common. The minimum is 0~fs, the maximum 36.5~fs. The mean and standard deviation are 0.36~fs and 1.15~fs, respectively. 
    Given the average transition time of 53.8~fs, transition trajectories share only 0.67\% of their snapshots. If we use the shared trajectory segments as a measure of correlation, then on average over 99\% are uncorrelated. 
    Note that in Fig.~\ref{fig:transitiontrajectories}e, the $y$-axis is scaled logarithmically to visualize the wide range of bin counts.
    
    % AVERAGE COMMITTOR FUNCTION
    \subsection{Averaged Committor Analyses}
    The committor of a configuration is the probability that a trajectory initiated from it will reach the product region $B$ before reaching the reactant region $A$.
    Configurations close to $A$ have committor values near zero, those close to $B$ near unity, and the transition state corresponds to a committor of $0.5$.
    For our NAFFS simulation, we can compute a closely related quantity, which we refer to as the \textit{averaged committor} $P_A(B\vert \lambda_j)$, defined as
    \begin{equation}
        P_A(B\vert \lambda_j)=\prod_{i=j}^n P_A (\lambda_{i+1}\vert \lambda_{i})
    \end{equation}
    The averaged committor is the probability that a trajectory initiated in $A$ that has already crossed interface $\lambda_j$ goes on to reach the product region $B$ (that is, eventually crosses the last interface, $\lambda_{n+1}=\lambda_B$). Hence, $P_A(B\vert \lambda_j)$ can be interpreted as a history-conditioned committor to $B$, averaged over the corresponding interface ensemble.

    \begin{figure}
    	\includegraphics[width=0.8\textwidth]{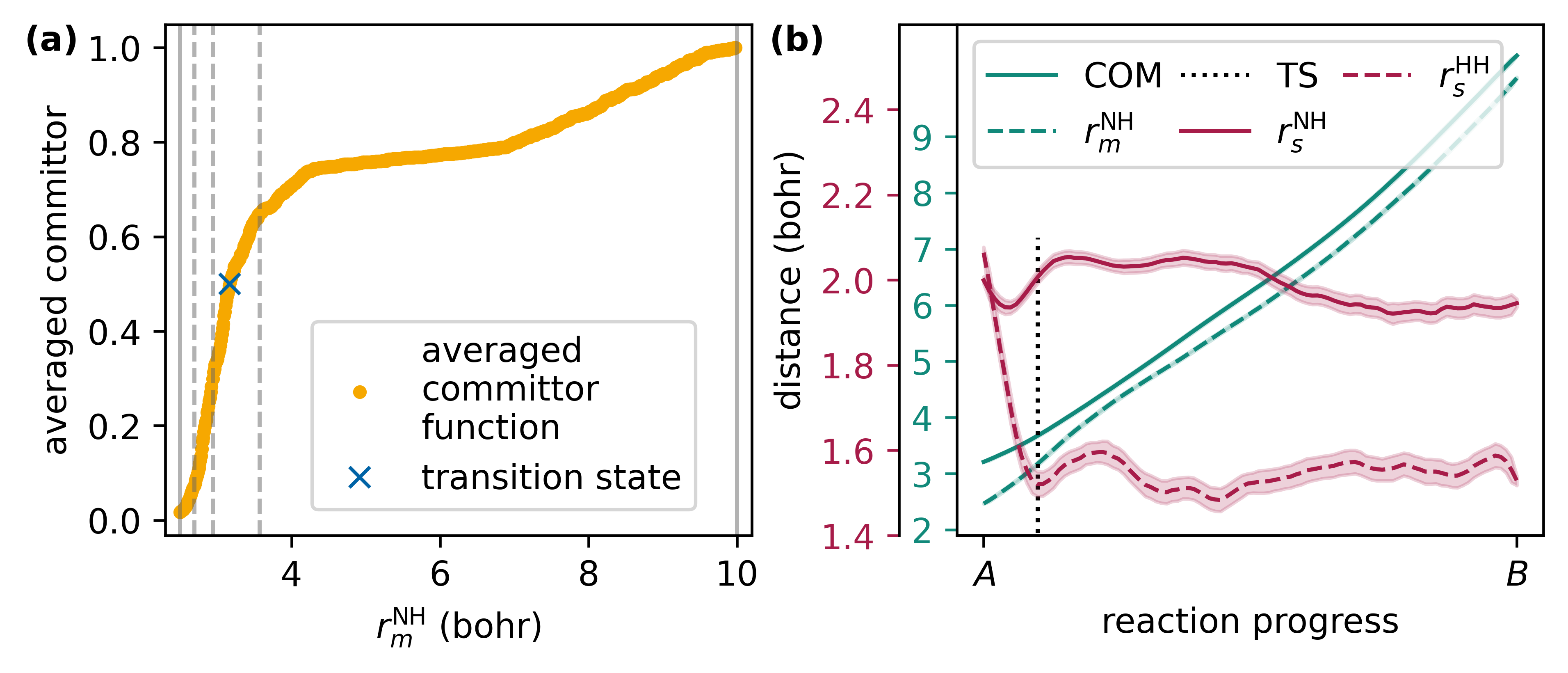}
    	\caption{a) Averaged committor function along the $r_m^{\ce{NH}}$ coordinate. The transition state (committor\;$=0.5$) is marked in blue.
        Initial and final region boundaries are shown as gray lines, interfaces used in the simulation as dashed lines. 
        b) Mean distances (in terms of $r_s^{\ce{NH}}$, $r_s^{\ce{HH}}$, $r_m^{\ce{NH}}$ and the center-of-mass distance (COM) between \ce{NH} and \ce{H2}) of the transition trajectories along the reaction progress (\textit{i.e.}, all paths are rescaled to unit length in time and shifted to a common time origin). 
        The respective standard deviations are shown as confidence bands, and the transition state (TS) according to the averaged committor is marked as dotted vertical line. 
        Two $y$-axes are used to improve visibility; the color of the curves corresponds to the color of the axis ticks they belong to.}
    	\label{fig:analyses}
    \end{figure}
    
    From the accepted and rejected NAFFS trial trajectories, the averaged committor can be evaluated on arbitrary interfaces $\left\lbrace\lambda_j\right\rbrace$, independently of the interface placement used in the NAFFS simulation itself.
    Figure~\ref{fig:analyses}a shows the averaged committor along $r_m^{\ce{NH}}$.
    It increases monotonically with $r_m^{\ce{NH}}$ and displays an extended plateau at $p_A\approx 0.75$ for $r_m^{\ce{NH}}$ between $5$ and $7$\,a$_0$.
    This plateau is consistent with Fig.~\ref{fig:transitiontrajectories}a, where several trajectories oscillate for extended periods, even up to 250\,fs, in this range of $r_m^{\ce{NH}}$ before reaching $B$, giving the impression that the system is briefly trapped.
    Inspection of the corresponding full geometries reveals that the \ce{H2} fragment orbits the \ce{NH} fragment, while both intramolecular bonds oscillate and occasionally rotate about their own axes; a behavior similar to roaming.\cite{Endo2020}

    The effective transition state, defined by an averaged committor value of $0.5$, is located at ${r_m^{\ce{NH}}}^{\ddagger}=3.16$\,a$_0$.
    It lies between the NAFFS interfaces $\lambda_2$ and $\lambda_3$ and is indicated by a blue cross in Fig.~\ref{fig:analyses}a.
    Trajectories starting in $A$ that cross the corresponding transition state interface $\lambda^\ddagger$ at ${r_m^{\ce{NH}}}^{\ddagger}$ have, by definition, a 50\% probability of proceeding to $B$.
    We emphasize that the distribution of points sampled on $\lambda^\ddagger$ is not the equilibrium distribution on that surface, since our ensemble is restricted to trajectories originating from $A$; however, this restriction is the physically relevant one for the light-induced reaction $A \rightarrow B$ studied here.
    On $\lambda^\ddagger$, the center-of-mass distance between \ce{NH} and \ce{H2} is $(3.67\pm 0.25)$\,a$_0$, and the distance of \ce{N} to the plane defined by the three hydrogens is $(0.80\pm0.46)$\,a$_0$.
    The corresponding interatomic distances are ${r_s^{\ce{NH}}}^{\ddagger}=(1.99 \pm 0.16)$\,a$_0$, ${r_m^{\ce{NH}}}^{\ddagger}=(3.16 \pm 0.03)$\,a$_0$, ${r_l^{\ce{NH}}}^{\ddagger}=(4.28 \pm 0.38)$\,a$_0$, ${r_s^{\ce{HH}}}^{\ddagger}=(1.53 \pm 0.15)$\,a$_0$, ${r_m^{\ce{HH}}}^{\ddagger}=(3.45 \pm 0.74)$\,a$_0$, and ${r_l^{\ce{HH}}}^{\ddagger}=(4.49 \pm 0.94)$\,a$_0$ (reporting means and standard deviations).
    Of the transition-state configurations, 65.5\% are located on S$_1$, 26.3\% on S$_0$, and 8.2\% on T$_1$.

    Figure~\ref{fig:analyses}b shows the evolution of four collective variables ($r_m^{\ce{NH}}$, $r_s^{\ce{NH}}$, $r_s^{\ce{HH}}$, and the center-of-mass distance between \ce{NH} and \ce{H2}) along the reaction progress with the corresponding standard deviations indicated as shaded bands.
    The reaction progress of a configuration is computed as the ratio of its frame number \textit{vs.} the total number of frames of the transition trajectory it belongs to. 
    The location of the effective transition state is marked by a dashed vertical line, which marks the point along the reaction progress when the average $r_m^{\ce{NH}}$ (green dashed line) has reached the transition state surface identified in Fig.~\ref{fig:analyses}a.
    $r_s^{\ce{HH}}$ drops sharply towards the transition state, reaching a minimum at $\lambda^\ddagger$, then rises to a maximum and levels off, indicating the formation of the \ce{H-H} bond.
    The $r_s^{\ce{NH}}$ variable passes through a minimum before the effective transition state and exhibits a broad maximum thereafter before decreasing again.
    Both $r_m^{\ce{NH}}$ and the center-of-mass distance increase almost linearly with the reaction progress, which confirms the suitability of $r_m^{\ce{NH}}$ as a reaction coordinate.

    % populations
    \subsection{State Populations}
\begin{figure}
    	\includegraphics[width=0.8\textwidth]{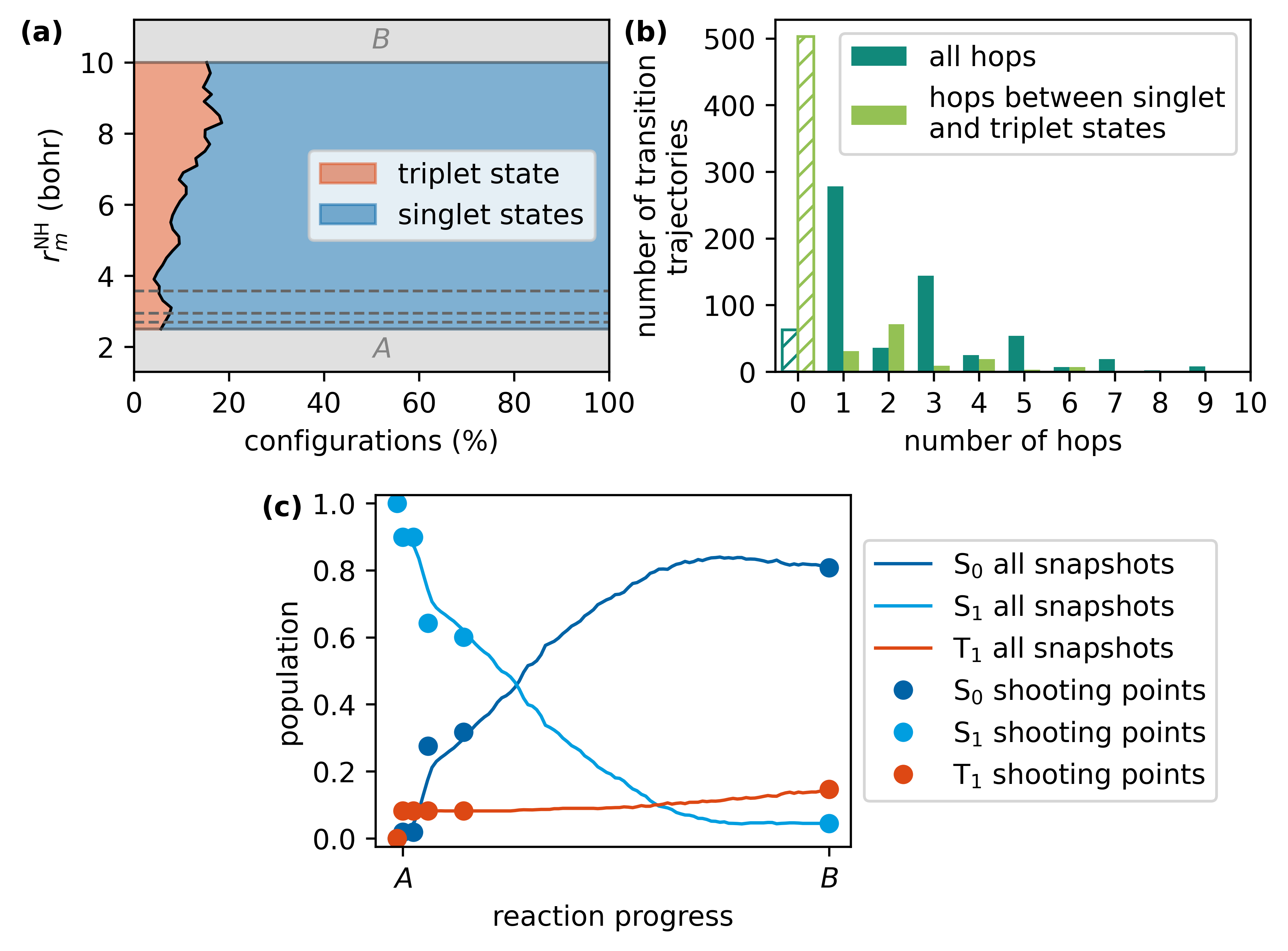}
    	\caption{Analysis of the populated states for the final transition trajectories. a) Percentage of time spent in singlet and triplet states along the reaction coordinate. b) Histogram of the number of hops, including both the combined total and the breakdown by singlet and triplet hops. Zero hops are highlighted by hatched bars. c) Population of the S$_0$, S$_1$ and T$_1$ state for all snapshots (solid lines) and only the shooting points at the interfaces (dots) along the reaction progress.}
    	\label{fig:state_analyses}
    \end{figure}

    The electronic-state composition of the transition trajectories is summarized in Fig.~\ref{fig:state_analyses}.
    Fig.~\ref{fig:state_analyses}a shows the percentage of singlet (\ce{S0}, \ce{S1}) and triplet (\ce{T1}) configurations along $r_m^{\rm NH}$.
    Although the singlet states dominate over most of the reaction, the triplet state plays a non-negligible role, accounting on average for 10.9\% of the configurations, with a maximum of 18.5\% at $r_m^{\rm NH}\approx 8.2$\,a$_0$ and a minimum of 4.2\% at $r_m^{\rm NH}\approx 3.8$\,a$_0$.
    The triplet population grows towards $B$ but remains a minority throughout the reaction.
    This is consistent with previous computational studies: Bachmair \textit{et al.}\cite{Bachmair2025} reported intersystem crossing in 10\% of their transition trajectories.

    Distributions of the number of hops per transition trajectory are shown in Fig.~\ref{fig:state_analyses}b.
    63 trajectories (9.8\%) show no hops at all and thus dissociate adiabatically on a single potential energy surface.
    The mean total number of hops per trajectory is 2.4 with a standard deviation of 2.2, and the maximum for a single trajectory is 15 hops.
    Restricting the analysis to hops between different multiplicities, the mean is 0.6 with a standard deviation of 1.3, and a maximum of 14. 
    143 trajectories (22\%) show at least one intersystem crossing, though some of these triplet excursions last only a few time steps and are not mechanistically relevant.
    The remaining 68\% of trajectories undergo only internal conversion between the singlet states.

    Fig.~\ref{fig:state_analyses}d shows the populations of the three electronic states along the reaction progress, where each transition trajectory is rescaled to unit length.
    All trajectories start on S$_1$ in $A$, and at the boundary of $B$ the populations are 80.7\% on S$_0$, 4.5\% on S$_1$, and 14.8\% on T$_1$. Since the generation of points on $B$ is conditioned in a way that they have to come from trajectories that started in $A$ and underwent the rare reaction, their probability distribution does not resemble the equilibrium distribution in $B$ one might expect based on Fig.~\ref{fig:theory}b, where the triplet is the lowest state for $B$ and S$_0$ and S$_1$ are basically degenerate.
    The populations sampled at the interfaces (dots in Fig.~\ref{fig:state_analyses}c) closely track those of the full ensemble.

    \subsection{Individual Transition Paths}
    
    \begin{figure}
    	\includegraphics[width=0.6\textwidth]{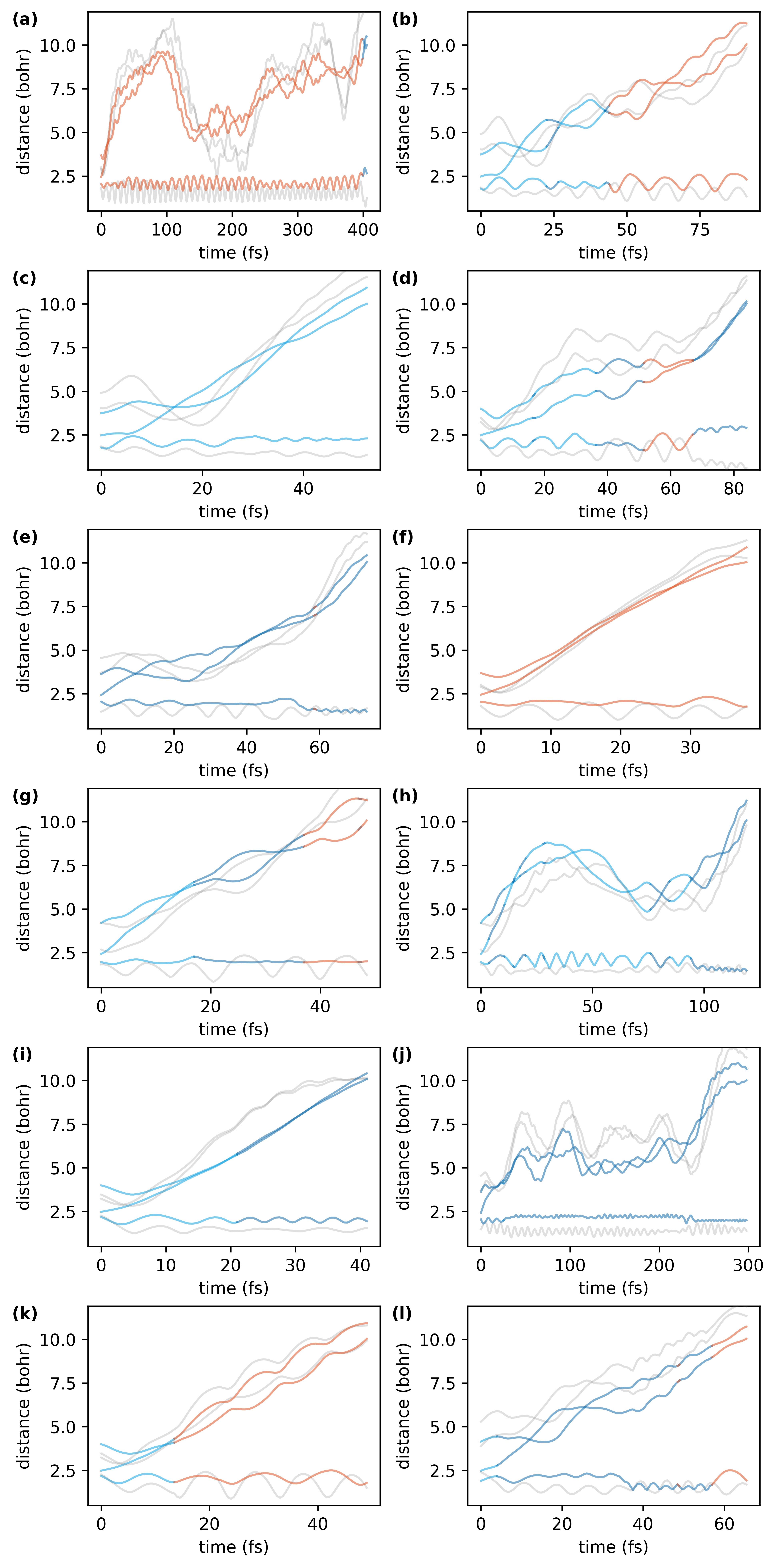}
    	\caption{Intramolecular distances \textit{vs.} time for exemplarily chosen transition trajectories (a)-(l). \ce{N-H} distances are shown in color, where the color indicates the state of the system (dark blue: S$_0$, light blue: S$_1$, red: T$_1$). \ce{H-H} distances are shown in gray.}
    	\label{fig:individualtransitions}
    \end{figure}

    % qualitatively categorize reactive paths
    Representative individual trajectories are shown in Fig.~\ref{fig:individualtransitions}a--l in terms of their \ce{N-H} (colored) and \ce{H-H} (gray) distances, and qualitatively agree with the trajectories reported by Wang \textit{et al.}\cite{Wang2022} and Bachmair \textit{et al.}\cite{Bachmair2025}
    The selection covers purely adiabatic trajectories on S$_1$ (panel c) and T$_1$ (panel f), trajectories featuring intersystem crossing (panels a, b, d, g, k, l), and trajectories undergoing only internal conversion (panels e, h, i, j).
    The transition trajectories can be qualitatively divided into three categories: (i) dissociations where the \ce{H2} is ballistically shot away (e.g., panels f and i), (ii) dissociations where a significant return of the \ce{H2} towards \ce{NH} happens (e.g., panels a and h), and (iii) dissociations where some oscillations without full return are visible (e.g., panels g and l). 271 transition trajectories (42\%) fall in category (i), 25 (4\%) in category (ii), and 350 (54\%) in category (iii). Hence, only for a minority (4\%) of reactive paths, the \ce{H2} significantly returns towards \ce{NH} during the dissociation, where we define ``significant'' as occurring when the $r_l^{\ce{NH}}$ distance becomes smaller than the $r_m^{\ce{NH}}$ distance recorded at a previous point in time for the respective trajectory. Although Fig.~\ref{fig:transitiontrajectories}a may give the impression that most trajectories undergo ballistic dissociation, this is a visualization artifact of plotting a large number of overlapping trajectories, which obscures  the behavior of individual paths.  
    In fact, 42\% of reactive paths dissociate ballistically, whereas the majority (58\%) exhibits oscillatory behavior of \ce{H2} with respect to \ce{NH}---these are also the trajectories that lead to the plateau in the averaged committor shown in Fig.~\ref{fig:analyses}a.

    %%%%%%%%%%%%%%%%%%%%%%%%%%%%%%%%%%%%%%%%%%%%%%%%%%%%%%%%%%%%%%%%%%%%%%%%%%%%%%%%%%%%%%%%%%%%%
    %\section{Discussion}
    \subsection{Computational Speedup}
    %\label{sec:speedup}

    From the thermostat-free reference trajectories of Bachmair \textit{et al.}\cite{Bachmair2025}, we can compute the average simulation time required to generate one rare transition trajectory in a brute-force setting.
    We consider two variants: in the \textit{naive} approach, a fixed number of trajectories is propagated for a fixed time and analyzed afterwards; in the \textit{optimal} approach, trajectories are monitored on-the-fly and terminated as soon as the product is formed.
    The optimal approach represents the best one can do with brute-force surface hopping, and an approximation of it has been used in previous work.\cite{Bachmair2025}
    Table~\ref{tab:speedup} compares the corresponding times to those obtained with NAFFS.

    NAFFS requires only 1723~fs of simulation time per transition trajectory on average, corresponding to a speedup of $106.3\times$ over the naive brute-force approach and $31.5\times$ over the optimal one. 
    If we take into account that NAFFS produces correlated transition trajectories (see Sec.~\ref{subsec:FluxEvents_TrajCorrelation} for a discussion), the simulation length for simulating an uncorrelated transition trajectory with NAFFS is 1736~fs on average, hence roughly the same as given in Table~\ref{tab:speedup}.
    In practical terms, a calculation that takes a month with the best previously available method can be reproduced by NAFFS in roughly one day, opening access to rare nonadiabatic reactions whose study  is computationally extremely demanding.

    \begin{table}[tb]
        \centering
      \caption{Average simulation time required to obtain one transition trajectory and respective speedup factors with respect to naive and optimal (opt.) brute-force simulations.}
      \begin{tabular}{c c c c}
        \hline
         & naive brute-force & optimal brute-force & NAFFS \\\hline
         simulation time (fs) & 183,150 & 54,198 & 1723 \\
         speedup factor naive & 1.0 & 3.4 & 106.3 \\
         speedup factor opt. & 0.3 & 1.0 & 31.5 \\\hline
      \end{tabular}
      \label{tab:speedup}
    \end{table}

    % rate constant
    \subsection{Reaction Rate Constant}
    The reaction rate constant from $A$ to $B$ obtained from the NAFFS method (Eq.~\ref{eq:rate}) is
    \begin{equation}
        \label{eq:rateNHH2}
        k_{AB} = (18.7 \pm 2.0)~{\rm ns}^{-1},
    \end{equation}
    which is in excellent agreement with the values previously calculated from brute-force surface hopping,\cite{Wang2022,Bachmair2025} 
    \begin{equation}
        k_{AB}^{\rm Wang} = (18.0 \pm 0.8)~{\rm ns}^{-1}
    \end{equation}
    and
    \begin{equation}
        k_{AB}^{\rm Bachmair} = (18.5 \pm 1.2)~{\rm ns}^{-1}
    \end{equation}
 
    (see Sec.~S4 for a detailed comparison).
    As with conventional surface hopping, the present implementation employs classical nuclear dynamics and therefore neglects nuclear quantum effects such as tunneling. Consequently, the rates reported here should be interpreted within the accuracy of the underlying dynamical approximation.
    A comparison of the rare \ce{NH + H2} channel with the dominant \ce{NH2 + H} channel, which is also observed in our flux simulation, is provided in Sec.~S4.
    
    Unlike the literature reference calculations,\cite{Wang2022,Bachmair2025} our NAFFS simulation uses a Langevin thermostat to provide the stochastic dynamics required by FFS-type methods.
    The thermostat parameters were chosen to reproduce the thermostat-free behavior of the system (Sec.~\ref{sec:computational}) while still yielding sufficiently uncorrelated transition trajectories.
    When using thermostats in gas phase calculations, special care must be taken and a comparison with non-thermostat results should be made to verify the suitability of the thermostat parameters, as we have done here.

    The dissociation timescale of $\sim 0.053$~ns associated with $k_{AB}$ is six orders of magnitude larger than the integration time step of 0.5~fs, illustrating the rare event regime that NAFFS allows us to access with a reasonable computational budget compared to traditional brute-force surface hopping simulations.

%%%%%%%%%%%%%%%%%%%%%%%%%%%%%%%%%%%%%%%%%%%%%%%%%%%%%%%%%%%%%%%%%%%%%%%%%%%%%%%%%%%%%%%%%%%%%
%%%%%%%%%%%%%%%%%%%%%%%%%%%%%%%%%%%%%%%%%%%%%%%%%%%%%%%%%%%%%%%%%%%%%%%%%%%%%%%%%%%%%%%%%%%%%
    \section{Conclusions}
    \label{sec:conclusions}

    This work establishes nonadiabatic forward-flux sampling (NAFFS) as a practical rare-event sampling method for full-dimensional molecular photochemistry, beyond the low-dimensional analytical models for which it had previously been validated.
    Specifically, we applied NAFFS to the rare molecular photodissociation channel of ammonia, \ce{NH3 + h\nu \rightarrow NH + H2}, using surface hopping dynamics on neural network potentials\cite{Wang2021a} for the two lowest singlet and the lowest triplet state.

    The sampled transition trajectories agree both qualitatively and quantitatively with the brute-force results of previous studies.\cite{Wang2022,Bachmair2025} 
    The committor analysis revealed a characteristic intermediate region in which a nascent \ce{H2} molecule remains in close proximity to the NH fragment. 
    Configurations in this region have approximately 25\% probability of recombining to form \ce{NH3}, indicating that the \ce{NH}$\cdots$\ce{H2} configuration is not merely a late-stage product geometry, but an intermediate region of the dissociation pathway. 
    This interpretation is further supported by the trajectory dynamics: although a substantial fraction of reactive paths dissociate ballistically, the majority exhibit oscillatory motion of \ce{H2} relative to \ce{NH} before final separation. 
    The rate constant obtained directly from the flux and conditional crossing probabilities, $(18.7 \pm 2.0)~{\rm ns}^{-1}$, is in close agreement with previous brute-force surface hopping simulations,\cite{Wang2022,Bachmair2025} validating NAFFS.  
    
    The reaction proceeds predominantly on the singlet states, with the triplet state accounting on average for about 10\% of the configurations along a transition trajectory.
    Of the 646 reactive trajectories, 68\% undergo internal conversion only, 22\% involve at least one intersystem crossing event, and 9.8\% dissociate adiabatically on a single potential energy surface.
    Thus, purely adiabatic pathways constitute the least common mechanism.
    
    Compared to brute-force surface hopping, NAFFS reduces the simulation time required per transition trajectory by a factor of about $100$ relative to the naive approach and by a factor of about $30$ relative to the optimal variant, consistent with the speedups previously observed for analytical model systems.\cite{Reiner2023} 
    This advantage is expected to increase further for systems with higher barriers, where reactive trajectories become even rarer in brute-force surface hopping simulations, whereas FFS is not penalized exponentially by the small overall transition probability.

    Alternative nonadiabatic transition path sampling approaches that avoid the use of a thermostat have recently been proposed, but so far only for proof-of-concept model setups.\cite{Ghamari2026,Yang2026}
    Such schemes may be particularly attractive for rare gas-phase photoreactions involving multiple electronic states, where the need to choose thermostat parameters is avoided.
    NAFFS, on the other hand, is naturally well suited to condensed-phase systems, in which thermostats are routinely employed.
    Combining NAFFS with the recently developed excited-state machine learning/molecular mechanics approach,\cite{Tiefenbacher2025} would, for example, enable the efficient study of rare nonadiabatic events in explicit solvent, opening the door to applications in complex condense-phase environments.

\begin{acknowledgement}
    This work is funded by the University of Vienna within the framework of the research platform ViRAPID. M.M.R. appreciates the support from the Vienna Doctoral School in Physics (VDSP). 
    The authors acknowledge the Austrian Scientific Cluster (ASC) for the generous allocation of computational resources. 
    Furthermore, the authors thank Brigitta Bachmair for insightful discussions.
\end{acknowledgement}

\begin{suppinfo}
    Code and data availability, additional implementation details, NAFFS calculation details of the partial trajectories for each interface, notes on the statistical convergence of the obtained results with respect to the size of the NAFFS simulation, considerations regarding the rate constant as a function of the friction parameter of the Langevin thermostat, additional details on the literature rate constants, and analysis of the frequent radical \ce{NH2 + H} photodissociation channel.
    The Supporting Information cited additional references.\cite{Granucci2012}
    The Supporting Information is available free of charge via the Internet at http://pubs.acs.org.

    The NAFFS code used to simulate all data in this paper has been published at \url{https://github.com/MadlenReiner/openpathsampling/tree/virapid}.
\end{suppinfo}

\section*{Author Contributions}
% author contributions according to CReDIT system
\textbf{M.M.R.}: Methodology, Software, Validation, Formal Analysis, Investigation, Data Curation, Writing - Original Draft, Visualization. \textbf{J.C.B.D.}: Methodology, Validation, Writing - Original Draft, Supervision. \textbf{L.G.}: Conceptualization, Resources, Writing - Review \& Editing, Supervision, Project administration, Funding acquisition. \textbf{C.D.}: Conceptualization, Methodology, Validation, Resources, Writing - Review \& Editing, Supervision, Project administration, Funding acquisition.

\bibliography{library.bib}

\includepdf[pages={-}]{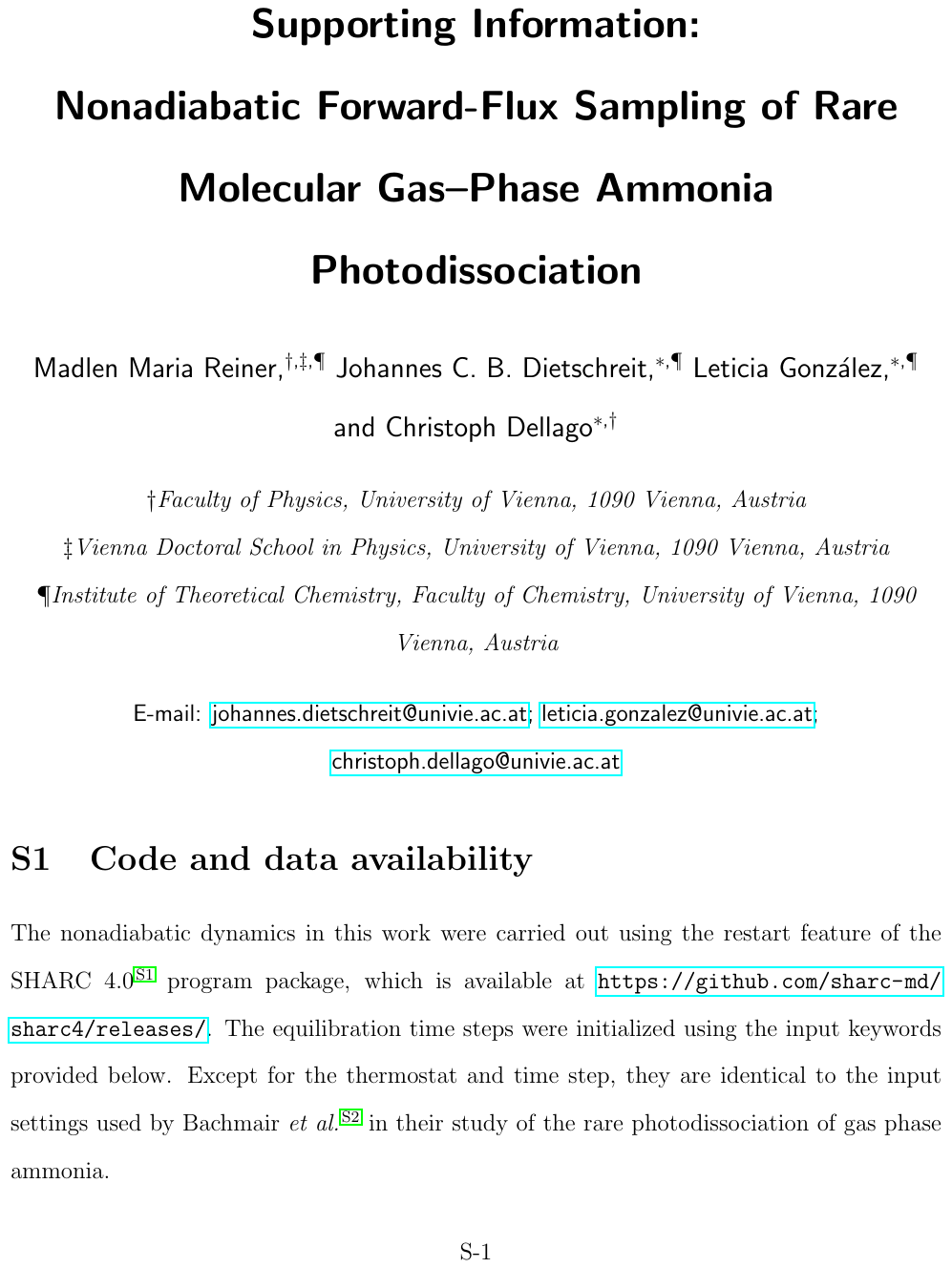}

\end{document}